\documentclass[10pt,preprintnumbers,twocolumn,amsmath,amssymb,nofootinbib,superscriptaddress]{revtex4-1}
\usepackage[latin1]{inputenc}
\usepackage{slashed}
\usepackage{amsmath}
\usepackage{textcomp}
\usepackage{amssymb}
\usepackage{amsfonts}
\usepackage{indentfirst}
\usepackage{color}
\usepackage{hyperref}
\usepackage[top=2cm,bottom=2cm,left=3cm,right=2cm]{geometry}

\newcommand{\be}{\begin{equation}}
\newcommand{\ee}{\end{equation}}
\newcommand{\bea}{\begin{eqnarray}}
\newcommand{\eea}{\end{eqnarray}}

\newcommand{\nn}{\nonumber\\}

\usepackage{graphicx,graphics}
\graphicspath{{figures/}}

\begin{document}

\hfill{KCL-PH-TH/2019-70}

\title{Saving the universe with finite volume effects}
\author{Jean Alexandre}  
%\email{jean.alexandre@kcl.ac.uk}
\affiliation{Theoretical Particle Physics and Cosmology, King's College London, Strand, London WC2R 2LS, UK}
\author{Katy Clough}
%\email{katy.clough@physics.ox.ac.uk}
\affiliation{Astrophysics, University of Oxford, DWB, Keble Road, Oxford OX1 3RH, UK}
 
\begin{abstract}
Setting aside anthropic arguments, there is no reason why the universe should initially favour a net expanding phase rather than one experiencing a net contraction. 
However, a collapsing universe containing ``normal'' matter will end at a singularity in a finite time. 
We point out that there is a mechanism, derived from non-perturbative effects in Quantum Field Theory in a finite volume, which may provide a bias towards 
expansion when the spacetime volume shrinks, by dynamically violating the null energy condition, without the need for modified gravity or exotic matter. 
We describe a scalar field component subjected to this non-perturbative effect in a cosmological background
and consider its impact on a contracting phase. 
We discuss how this could dynamically generate the necessary initial conditions for inflation to get started, or form part of the mechanism for a non-singular cosmological bounce.
\end{abstract}

\maketitle

\section{Introduction}

The Hawking-Penrose theorems \cite{Hawking:1969sw, Borde:1993xh} imply that in the absence of matter which breaks the null energy condition (NEC) $\rho + p \ge 0 ~ $,
the universe, run backwards, must begin at a singularity.
A corollary of this is that, under generic conditions, a collapsing universe cannot transition to a growing one, and thus if the universe started in a collapsing state, it should have ended. 
This statement can also be formulated in terms of the ADM decomposition \cite{barrowandtippler, Arnowitt:1960es, Kleban:2016sqm}, and forms the basis for work on cosmological bounces 
\cite{Khoury:2001bz,Steinhardt:2001st,Khoury:2001wf}.

Such ideas are of general interest in the context of early universe cosmology, since the Universe could in principle have started in either an expanding or contracting phase. 
Whilst one can make an anthropic argument that contracting universes are simply not conducive to life, it is interesting to explore whether there are deeper, 
more fundamental reasons for our universe to favour expansion.

Consider the dominant early universe paradigm of inflation, which solves the homogeneity and isotropy problems of the universe by postulating a 
period of accelerated expansion  \cite{Guth:1980zm,Linde:1981mu,Albrecht:1982wi,Starobinsky:1980te}, whilst also providing a scale invariant spectrum of initial perturbations. 
The question of how inflation {\it actually got started} is not yet fully resolved. In particular, if inflation requires ``special'' initial conditions in order to proceed, 
at least one of its key motivations may be undermined.

%%%%%%%%%%%%%%%%%%%%%%%%%%%%%%%%%%%%%%%%%%%%%%%%%%%%%%%%%%%%%%
\begin{figure}[t]
	\centering
	\includegraphics[scale=0.5]{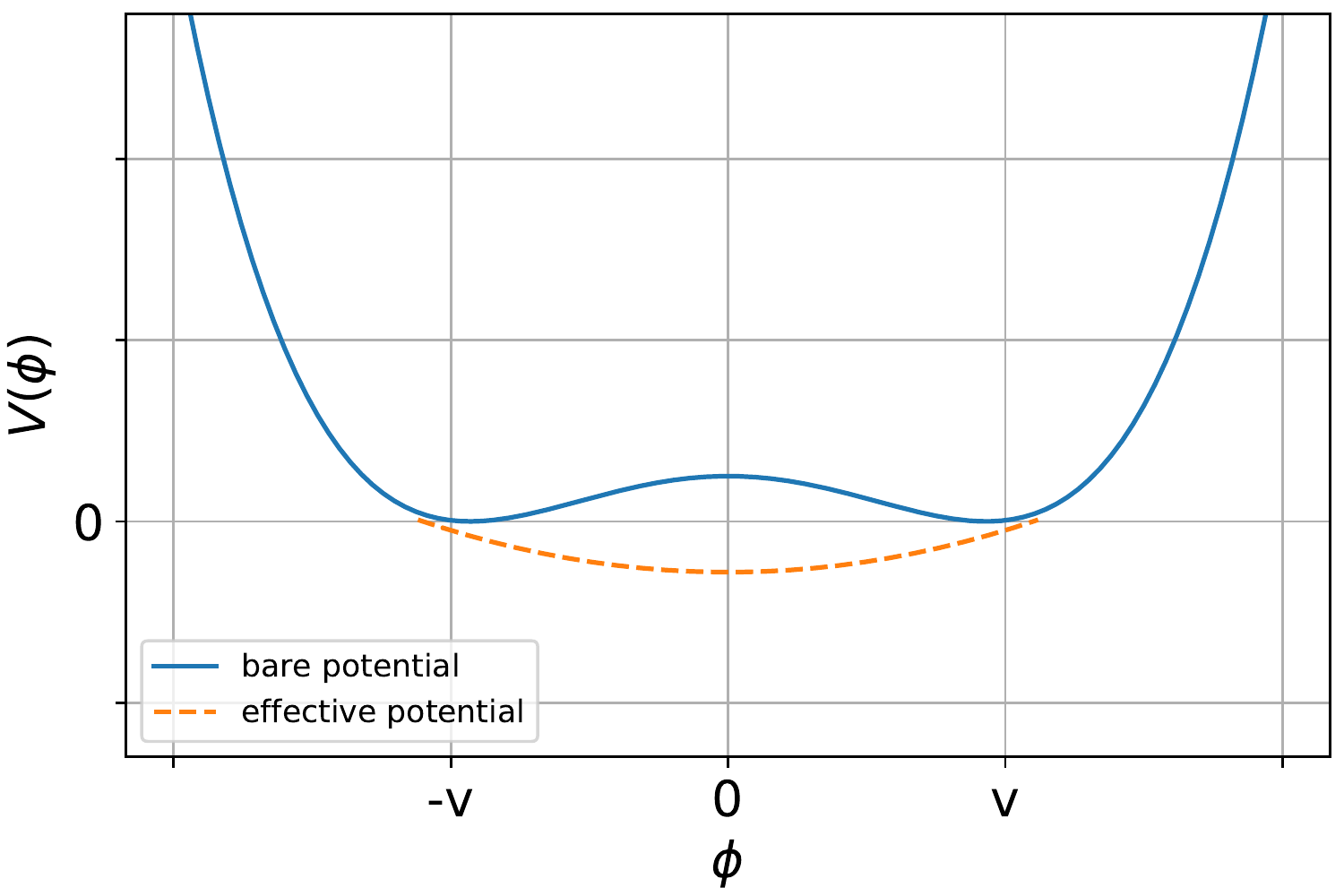}
	\caption{Illustration of finite volume effects on a concave bare potential. Symmetry is restored by tunnelling between the two bare minima, 
	which results in a negative energy density in the ground state.}
	\label{Fig:Potential}
\end{figure}
%%%%%%%%%%%%%%%%%%%%%%%%%%%%%%%%%%%%%%%%%%%%%%%%%%%%%%%%%%%%%%

Numerical and analytic studies have considered the question of what initial conditions can be tolerated in a variety of different models (see~\cite{Brandenberger:2016uzh} for a short review). 
In particular, recent numerical work showed that large field models are strongly robust to initial inhomogeneities in the inflaton field \cite{East:2015ggf}, 
whilst small field models are more sensitive to horizon sized perturbations \cite{Clough:2016ymm, Clough:2017efm}. 
Yet these studies are open to the criticism that they bias their results by assuming that the universe begins in an expanding phase. (In \cite{Clough:2017efm}, 
a mixture of expanding and contracting phases were considered, but the average expansion rate was positive.) 
This is a choice that must be made in setting up the initial conditions, as the first Friedmann equation is agnostic to the initial sign of $\dot{a}$. 
In an initially contracting universe, invoking only standard General Relativity (GR) and single field slow roll, inflation will fail, regardless of the chosen model.

In the alternative paradigm of bouncing cosmologies, the smoothing of the Universe may be achieved in a contracting phase dominated by a form of matter with equation of state parameter 
$w \gg 1$ in Ekpyrotic models \cite{Khoury:2001bz,Steinhardt:2001st,Khoury:2001wf} or $w=1$ in the Pre-Big-Bang scenario \cite{Gasperini:1992em}. 
The generation of a scale invariant spectrum of perturbations is less easy to achieve, with most models invoking special mechanisms, 
with the exception of the Matter Bounce scenario \cite{Finelli:2001sr} in which the contraction is dominated by cold matter. 
Clear and up to date reviews of the topic are given in \cite{Ijjas:2018qbo}, and \cite{Brandenberger:2016vhg}. 
However, whilst several suggestions have been made \cite{bouncemechanism}, the origin of the cosmological bounce, that leads to our current expansion, is still an open question.

The requirement to break the NEC is generally considered rather exotic, apart from the well-known Casimir effect.
However, consider the case where one has a secondary field which sources a negative energy density, 
and furthermore where the magnitude of the negative energy density {\it grows} as the spacetime volume shrinks, as in the Casimir effect. 
In the expanding case, the effect soon dilutes away. In the collapsing case, the negative energy density increases and can balance the other energy contributions, 
slowing and eventually ending the collapse at a finite size. 
If, in addition, the effective equation of state violates the NEC during the collapse, it provides the necessary push for the spacetime to transition to expansion. 

In this paper we point out that such an effect is {\it exactly} that expected due to finite volume effects in a scalar field which is subject to a double-well bare potential,
as illustrated in Fig. \ref{Fig:Potential}. Assuming the volume is smaller than some typical volume, related to the parameters of the bare potential, 
quantum fluctuations can tunnel between the degenerate vacuua. As a consequence, the vacuum expectation value of the field vanishes and symmetry is restored. 
If the bare vacuua have vanishing energy density, the necessity for
convexity of the effective potential implies that quantum fluctuations dynamically generate a negative energy density in the ground state.
The result can be generalised to higher-dimensional field spaces with a continuous set of degenerate vacuua - a Higgs-like Mexican hat potential, for example - as was shown in \cite{Tsapalis}
in a flat spacetime. 

In this article, we explain why, in the context of a spatially flat Friedman-Lema\^itre-Robertson-Walker (FLRW) background, 
the magnitude of the energy density which is generated is proportional to the third inverse power of the scale factor $a^{-3}$.
We also find that our ``convexion'' field behaves as a pressureless fluid because of the non-trivial volume-dependence 
of the effective action, and thus has equation of state $w=0$. Since it has a \emph{negative} energy density, it thus violates the NEC. 

We finish by describing a toy model in which we combine a cosmological constant with our ``convexion'' field, in a situation where it sees the above non-perturbative convexity effect. 
This illustrates very simply the bias towards expansion. 
Whilst such a result is known and expected for any NEC violating fluid, we emphasise that our model contains \emph{no new physics} - no non standard couplings or modifications to 
the gravity sector. It is a natural dynamical consequence of the bare potential of the scalar, evolving in a cosmological spacetime.

\section{Finite volume and the effective potential in FLRW spacetime}

To define the volume in a cosmological spacetime with FLRW metric
\be\label{FLRW}
ds^2 = -dt^2 + a^2(t) (dx^2 + dy^2 + dz^2) ~,
\ee
we assume that the flat 3-dimensional space is a 3-torus with volume $V_0$, 
and is therefore periodic in comoving coordinates. The comoving volume is then 
\be
V^{(3)}(t)=V_0 a^3(t)~, 
\ee
and the 4-dimentional volume appearing in the action is 
\be
V^{(4)}=\int d^4x\sqrt{|g|}=V_0 \int_{t_i}^{t_f}dt~a^3(t)~,
\ee
where $T\equiv t_f-t_i$ is identified with a typical time for quantum fluctuations to propagate over the volume $V_0$.

\subsection{Condition for convexity}

Phase transitions are generally considered in the limit of ``infinite'' volume,  
which is required for the wave functions corresponding to different degenerate vacuua 
to be orthogonal and not overlap in field space \cite{Miransky}. In the case of the Higgs mechanism with vacuum $v\simeq240$ GeV, ``infinite'' volume means large compared to 
a scale proportional to $v^{-3}\simeq(10^{-18}m)^3$, which is obviously the case and thus ensures spontaneous symmetry breaking (SSB) in the Higgs model. 
For this reason, the one-particle irreducible (1PI) effective potential
of a model with SSB is derived from a partial partition function, based on one vacuum only, and not on the whole field space \cite{Plascencia:2015pga}.  

However, in the situation where the volume is finite, tunnelling between different vacuua is allowed and, as a consequence,  
the effective potential of a scalar theory must be convex \cite{convexity}: symmetry is restored by quantum fluctuations. 
Such a non-perturbative effect is possible only as the result of the competition between different 
saddle points in the partition function which defines the quantised theory \cite{competition}. 

We derive here the condition for tunnelling between vacuua to occur, and therefore the effective action to be convex.
For this we first estimate the width, in field space, of the distribution of fluctuations over the two different homogeneous 
saddle points in the partition function. We then require this width to be of the order of the separation of these two
saddle points, and thus derive a condition on the 4-dimensional volume for tunnelling to happen.

We are looking for an estimate only of the condition for this tunnel effect to occur, 
and for clarity we therefore restrict ourselves, in this specific section, to a situation where the scale factor changes slowly compared to typical quantum fluctuations, 
and where flat spacetime is a good approximation. This will be even more valid in the vicinity of a cosmological bounce of an FLRW spacetime, which is the regime we focus on.

We start with the bare potential
\be\label{sympot}
U_{bare}(\phi)=\frac{\lambda}{24}(\phi^2-v^2)^2~,
\ee
and consider the partition function for vanishing sources. We choose a vanishing energy for the vacuua in order to make explicit the dynamical generation of a 
negative energy in the true vacuum. This also avoids the choice of a fined-tunned vacuum energy, and could be motivated, for example, by an underlying supersymmetric model.
The equation of motion for the microscopic scalar field is
\be
\Box\phi+\frac{\lambda}{6}\phi(\phi^2-v^2)=0~, 
\ee
and $\phi$ is expanded above the vacuua $\pm v$ as
\bea\label{expandphi}
\phi_1(x)&=&v+\psi_1(x)\\
\phi_2(x)&=&-v+ \psi_2(x)~.\nonumber
\eea
The semi-classical approximation described below is based on classical configurations, such that we consider here on-shell fluctuations. These satisfy, for $(k=1,2)$,
\bea
\Box\psi_k+m^2\psi_k&=&{\cal O}(\lambda)\\
\mbox{with}~~~~m^2&\equiv&\frac{\lambda}{3}v^2~,\nonumber
\eea
and we consider $(\lambda,m^2)$ independent parameters, from which the vacuum $v$ is determined. With appropriate boundary conditions, we have then
\bea
\psi_k(x)&=&\xi_k\sin(q_\mu x^\mu)+{\cal O}(\lambda) \\
\mbox{ with}~~~~ q^2&=&m^2~,\nonumber
\eea
where $\xi_k$ are constant amplitudes to be integrated in the partition function. 
The terms ${\cal O}(\lambda)$ in the fluctuations $\psi_k$ corresponds to harmonics with momentum $2q_\mu$ and $3q_\mu$, such that
the action for these fluctuations is $(k=1,2)$
\bea
S_k&=&\frac{1}{2}\int d^4x\Big(q^2\xi_k^2\cos^2(q_\mu x^\mu)-m^2\xi_k^2\sin^2(q_\mu x^\nu)\nn
&&~~~~~~~~~-\frac{\lambda}{12}\xi_k^4\sin^4(q_\mu x^\mu)+{\cal O}(\lambda^2)\Big)\nn
&=&-\frac{\lambda}{64}\xi_k^4V^{(4)}+{\cal O}(\lambda^2)~.
\eea
In the partition function, the latter actions lead to distributions peaked over the vacuua $\pm v$, with half-width 
\be
\Delta=\left(\frac{64}{\lambda V^{(4)}}\right)^{1/4}~.
\ee
For these distributions to overlap considerably in field space, and therefore tunnelling between vacuua to occur, 
this half-width should be at least of order $v$, which leads to the convexity condition
\be\label{conditionV4}
\lambda v^4V^{(4)}\lesssim64~.
\ee
This bound is quite conservative - tunnelling would occur beyond it, but at a slower rate.

\subsection{Wilsonian approach}

The disappearance of the concave part of the potential has been studied in the Wilsonian context \cite{Wilsonian}, 
which in spirit is closer to the Effective Field Theory context than the 1PI potential, and therefore is more commonly used in the framework of cosmology.
The Wilsonian effective potential is flat between the bare vacuua (and therefore convex), which corresponds to the so-called Maxwell 
construction \cite{Maxwell}, by analogy with the transition from liquid to vapour, when studied with the Van de Waals equation of state. 
(Note that the Maxwell construction has been used to describe spinodal inflation \cite{spinodal}, which is different to the mechanism we describe here).
How is this consistent with symmetry restoration? In a flat effective potential, the field is equally likely to be found at any point in the flat section, 
and thus it describes the coexistence of different phases, each with different ground states, which {\it on average}
give a symmetric ground state. As the volume tends to infinity, tunnelling is suppressed, and bubbles of each phase nucleate, such that at any localised point 
in space one need only consider the potential around the relevant vacuum for that phase. It is in this large volume limit that one usually considers the potential for the inflaton, for example.
  
In this work, we consider the 1PI effective potential since, unlike the Wilsonian one, it takes into account finite volume effects and leads to symmetry restoration locally, not just on average.
The 1PI effective potential is identical to the Wilsonian effective potential for infinite volume only.

\subsection{Effective potential}

If we assume a regime where the volume satisfies the condition in Eq. (\ref{conditionV4}), 
the partition function is dominated by both saddles points corresponding to the two vacuua. The semi-classical approximation 
for the effective action is then described in \cite{Tsapalis} and we follow the same steps, also valid for curved space time, in the Appendix. 
The convex effective action we obtain is 
\bea\label{Seff}
S_{eff}[\phi_c]&=&-K+\frac{4A}{1+8A}\left(\frac{\phi_c}{v}\right)^2\\
&&+\frac{8A(-3+12A+128A^3)}{3(1+8A)^4}\left(\frac{\phi_c}{v}\right)^4\nn
&&+{\cal O}(\phi_c/v)^6~,\nonumber
\eea
where $K$ is a constant, $\phi_c$ is the homogeneous expectation value of the convexion, and 
\be
A\equiv\frac{\lambda}{24}v^4V^{(4)}~. \label{eqn:defA}
\ee
We note the following points
\begin{itemize}
\item Eq. (\ref{Seff}) is only valid for $|\phi_c|\ll v$. As explained in the Supplemental Material the semiclassical approximation for $|\phi_c|\gg v$ leads to an effective action
which is identical to the bare action.
\item Given the expression for $A$, one can see that the convexity condition Eq. (\ref{conditionV4}) is satisfied for $A\lesssim 8/3$; 
\item In addition to being convex, one can see that the effective action has a non-trivial volume-dependence, and is not extensive. The effective potential, 
obtained after dividing by $V^{(4)}$, therefore depends on the volume in a non-trivial way;
\item The constant $K$ is determined from continuity of the effective action (\ref{Seff}) at $\phi_c=\pm v$, which leads to
\be\label{K}
K\simeq\frac{4A}{1+8A}+\frac{8A}{3(1+8A)^4}(-3+12A+128A^3)~.
\ee
\end{itemize}
For $A\gtrsim1$, Eq. (\ref{K}) leads to
\be\label{Kbis}
K\simeq\frac{7}{12}-\frac{5}{48A}+\frac{11}{384A^2}+\cdots~.
\ee
In the large-$A$ limit, the resummation of all the powers of $\phi_c/v$ was calculated in \cite{Alex} and leads to
\bea\label{resummation}
S^{A\gg1}_{~eff}[\phi_c]&=&-\ln2+\frac{1}{2}\left(1-\frac{\phi_c}{v}\right)\ln\left(1-\frac{\phi_c}{v}\right)\nn
&&+\frac{1}{2}\left(1+\frac{\phi_c}{v}\right)\ln\left(1+\frac{\phi_c}{v}\right)~.
\eea
This expression is independent of $A$ and is therefore intensive. 
Although this feature is valid for large $A$, one can see from the Eq. (\ref{Kbis}) that 
$K$ is already approximately volume-independent if $A$ is of order 1, and therefore within the convexity regime described by Eq. (\ref{conditionV4}).

The effective action in this intensive regime should not depend on the scale factor $a(t)$, and its value in the vacuum can therefore be written
\be
S_{eff}[0]=-K=-\int d^4x\sqrt{|g|} ~\frac{\rho_0}{a^3(t)}~,
\ee
where $\rho_0$ is a positive constant. In the present semi-classical approximation we have
\be
\rho_0=\frac{K}{V_0 T}~ .
\ee
A more accurate value would require a better approximation for the effective action in the region $|\phi_c|\simeq v$. One expects then that $\rho_0$ depends on the 
parameters appearing in the bare potential, but at this stage we leave $\rho_0>0$ as a free parameter. The energy density $\rho_c$ 
and the pressure $p_c$ in the vacuum state $\phi_c=0$ are finally
\bea\label{rhop}
\rho_c&=&\left.\frac{2}{\sqrt{|g|}}~\frac{\delta S_{eff}}{\delta |g_{00}|}\right|_{\phi_c=0}=-\frac{\rho_0}{a^3(t)}~<0\\
p_c&=&\left.\frac{2}{\sqrt{|g|}}~\frac{\delta S_{eff}}{\delta |g_{11}|}\right|_{\phi_c=0}=0~,\nonumber
\eea
such that the NEC is violated in the ground state, since $p_c+\rho_c<0$. Note that this vacuum state consistently satisfies 
the continuity equation for a fluid $\dot\rho_c=-3H(\rho_c+p_c)$.

We finally note that the present effect is in principle possible for $1\lesssim A\lesssim 8/3$, which implies a restriction on the validity of the model.
Nevertheless, as already mentioned, the upper bound is quite conservative, and the overlap of wave functions corresponding to the different vacuua remains non-negligeable for 
larger values for $A$. We must assume though, for the present mechanism to be valid, that the Universe indeed goes through an appropriate regime, where convexion tunnelling between 
bare vacuua indeed occurs.

\subsection{Casimir effect}

We end this section with comments regarding the Casimir effect, which could potentially have a role for small volumes.
The study of Casimir effect in the context of Cosmology was introduced in \cite{Zeldovich:1984vk}, where a massless scalar field in a 3-torus has a repulsive Casimir effect,
and therefore contributes to an expansion of the Universe. More generally, the Casimir effect is either repulsive or attractive, depending on the coupling of the scalar
field to the metric, but also on the 3-space geometry/topology (see \cite{reviewCasimir} for a review). 
A massless scalar field conformally coupled to the metric, and experiencing the Casimir effect, has an energy density $\rho$ and a pressure $p$ which scale as 
\be\label{energyCasimir}
\rho\propto\frac{\pm1}{a^4(t)}~~~~\mbox{and}~~~~p=\frac{\rho}{3}~,
\ee
and which could potentially destabilise the tunnelling effect for small scale factor, since the convexion energy density scales as $a^{-3}(t)$. Nevertheless, 
for a massive scalar field, the energy density (\ref{energyCasimir}) is suppressed as \cite{reviewCasimir}
\be
|\rho|\propto\frac{(ml_0)^{5/2}}{a^{3/2}(t)}e^{-ml_0a(t)}~~~~\mbox{with}~~~~ml_0a(t)>>1~,
\ee
where $l_0^3\sim V_0$.
Hence one can expect that tunnel effects remain dominant, in a certain region of parameter space at least. 

The Casimir effect has been calculated for different couplings of the scalar field to the metric \cite{Herdeiro:2005zj},
on a 3-sphere though, and we note that quantum corrections arising from tunnel effect have always the same sign, independently of the geometry/topology of the 3-space.

\section{Cosmological illustration}

\begin{figure}[ht]
    \vspace{0.25cm}
	\centering
        \includegraphics[scale=0.26]{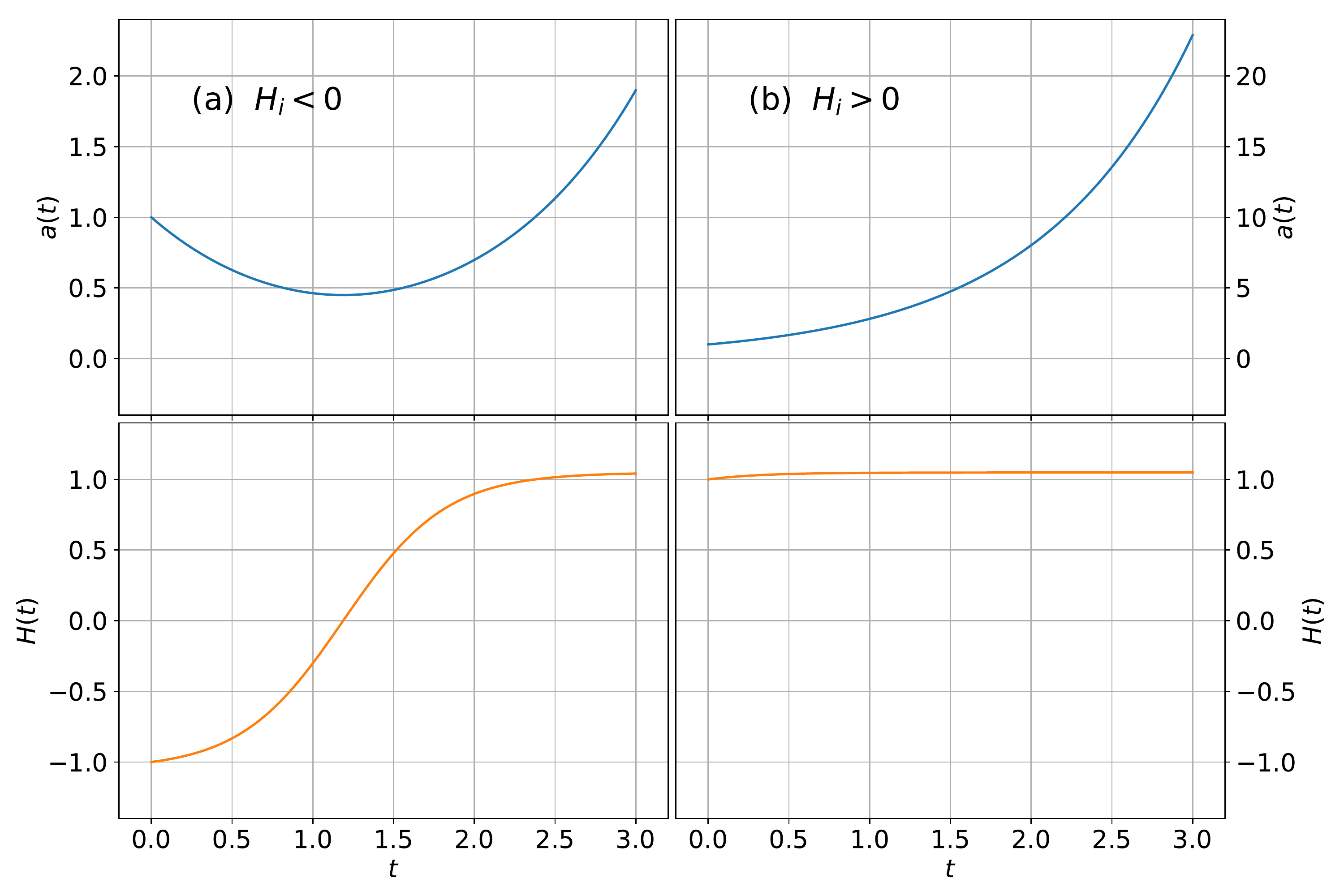}
	\caption{An illustration of the evolution of the scale factor $a$ and Hubble constant $H$ as a function of time $t$ for the cases of initial contraction 
        (a) and initial expansion (b) in our toy model. We see in the collapsing case that the convexion field provides the necessary bounce to transition to a period of exponential expansion. 
        If the initial condition is already expanding, the effect of the convexion is quickly diluted away.}\label{Fig:Evolution}
\end{figure}

We now describe a toy cosmological model as an illustration of how the above dynamical 
mechanism could work to favour expansion. We consider the FLRW metric Eq. (\ref{FLRW})
with Hubble constant $H=\dot a/a$, and we assume that the matter content is dominated by two components, 
both of which satisfy the continuity equation $\dot\rho=-3H(\rho+p)$ individually:
\begin{itemize}
    \item a cosmological constant, with energy density $\rho_f$ and pressure $p_f=-\rho_f$ (this could be a scalar field slow-rolling in a convex potential);
    \item the ``convexion'' field, with a vanishing expectation value $\phi_c=0$, and an energy density and pressure given by the expressions in Eq. (\ref{rhop}).
\end{itemize}
Using the Friedmann equations
\bea
H^2&=&\frac{\kappa}{3}(\rho_f+\rho_c)\\
\dot H+H^2&=&-\frac{\kappa}{6}(\rho_f+\rho_c+3p_f+3p_c)~,\nonumber
\eea
where $\kappa\equiv8\pi G$, the evolution equation is
\be\label{Friedmann2}
\dot H =\frac{\kappa}{2}\frac{\rho_0}{a^3} ~ > ~ 0~,
\ee
to be solved with initial values $a_i=1$, and $H_i$ which satisfies the constraint
\begin{equation}
\label{Friedmann1}
H_i^2 = \frac{\kappa}{3}\left(\rho_f - \rho_0\right) ~.
\end{equation}
The resulting evolution is shown in Fig. \ref{Fig:Evolution} for the cases of initially positive and negative $H$. 
It is clear from Eq. \eqref{Friedmann2} that for our chosen components the NEC is violated and thus the acceleration of the scale factor 
$\ddot a$ is positive, providing the necessary bounce to expansion in the contracting case.

%%%%%%%%%%%%%%%%%%%%%%%%%%%%%%%%%%%%%%%%%%%%%%%%%%%%%%%%%%%%%%
%%%%%%%%%%%%%%%%%%%%%%%%%%%%%%%%%%%%%%%%%%%%%%%%%%%%%%%%%%%%%%

\section{Conclusions}

In this article we have suggested a connection between finite volume effects in Quantum Field Theory and Early Universe Cosmology, to answer the question - {\it why is the universe expanding?} 
We have postulated the existence of a ``convexion'' field which sees different degenerate bare vacuua. Quantum tunnelling effects result in a pressureless fluid with negative energy density,
which plays a role at small volumes, but dilutes away during an expanding period. Such a fluid breaks the NEC, and thus provides a mechanism for a transition from contraction to an expanding phase.
Unlike other proposed mechanisms, this does not involve non standard couplings for the field or a modification of the gravity sector, and thus does not result in ghosts -
the dynamical mechanism is valid for any standard scalar field which can tunnel between degenerate vacuua.

One naturally asks about the regime in which this effect is valid. As discussed in the text, convexity, and thus a negative energy density, 
is achieved for any 4-dimensional volume small enough to satisfy the condition in Eq. (\ref{conditionV4}), which leads to a condition $A\lesssim 8/3$. 
However, an intensive effective action (which provides the $w=0$ equation of state), is valid for $A\gtrsim1$, which sets a minimum volume via Eq. (\ref{eqn:defA}). 
The mechanism we describe here is therefore valid within a regime where the volume is not ``too big'', but also not ``too small''. 
Of course, at some very small volume, quantum gravity effects should also play a role, and our description will break down.
Similar non-perturbative effects have been discussed in the works \cite{DiTucci:2019xcr,Bramberger:2019zks}, 
which take into account finite volume effects in a semiclassical description of quantum gravity. There the authors discuss 
the contribution of two different saddles points in the path integral for gravity, and they derive a {\it minimum} volume in order to decouple these saddle points and therefore 
have a well-defined cosmological evolution. 
For gravity to be classical, as it is in our work, the minimum volume we consider must be greater than the minimum volume defined in 
\cite{DiTucci:2019xcr,Bramberger:2019zks}. In this way quantisation of gravity would not affect the results in our regime of validity. 
Given that there are free parameters in both models, this inequality can easily be satisfied for a large range of parameter space.

Much work remains to be done to elucidate this interesting effect. Firstly, one must justify the interpretation of equilibrium field theory in a dynamical spacetime. 
One should also investigate what the convexity condition (\ref{conditionV4}) means for 
the separation of ultraviolet modes of the convexion from its infrared description in terms of a vacuum expectation value. 

A better approximation for the calculation of the partition function, including non-uniform saddle points
and quantum fluctuations above each saddle point, should also be made. This is quite an involved calculation, but we would expect any modifications 
to be relevant only away from the vacuum state which we consider here.

Finally, we have considered only the coexistence of the convexion with a cosmological constant. 
Matter or radiation like components would dominate over the convexion during a collapsing period, and in the present simplified description one must make the 
strong assumption that they are not present.
Further studies are necessary to understand the interplay between the convexion and different matter components, once kinetic terms are included. 
The latter might play an important role, depending on the flatness of the potential around the true minimum, and should allow us to understand better this interplay.

\vspace{1.0cm} 

{\it Acknowledgements:}
We are grateful to Eugene Lim for his insightful comments on an earlier draft. The work of JA is partially supported by STFC (grant ST/P000258/1).
KC acknowledges support from the European Research Council.

\section*{Appendix: Semi-classical approximation}

We derive here the effective action for the bare action 
\be
S[\phi]=\int d^4x\sqrt{|g|} \left(\frac{g_{\mu\nu}}{2}\partial^\mu\phi\partial^\nu\phi-U_{bare}(\phi)\right)~,
\ee
with
\be
U_{bare}(\phi)=\frac{\lambda}{24}(\phi^2-v^2)^2~,\nonumber
\ee
when both vacuua are taken into account in the partition function
\be
Z[j]=\int{\cal D}[\phi]\exp\left(-S[\phi]-\int d^4x\sqrt{|g|} j\phi\right)~.
\ee
The source $j$ is chosen constant, since we are interested in the effective potential only.
The semi-classical approximation corresponds to replacing the path integral by the sum over the relevant saddle points, 
and neglecting quantum fluctuations above the latter. 
Taking into account the source term, the bare vacuua are not degenerate, and the Coleman configuration, which asymptotically 
reaches the two different vacuua \cite{bounce}, 
should in principle play a role. Nevertheless, as we will show, the ground state of the system consists of a vanishing background field $\phi_c=0$, 
which corresponds to a vanishing source $j$, for which this configuration does not contribute. 
If we focus on the background field satisfying $|\phi_c|\ll v$, or equivalently $|j|\ll j_c$, it is therefore enough to consider uniform saddle-point configurations, 
and we follow here similar steps as those described in \cite{Tsapalis}. 

These uniform saddle points are solutions of the equation
\be\label{equamot}
\frac{dU_{bare}}{d\phi}+j=0~,
\ee
and depend on the amplitude of the source $j$:
\begin{itemize}
 
\item if $|j|>j_c$, where $j_c=\lambda v^3/(9\sqrt3)$, then Eq. (\ref{equamot}) has one real solution $\phi_0$;

\item if $|j|<j_c$, then eq.(\ref{equamot}) has three real solutions, one of which is a maximum and the two others are the local minima relevant for the partition function
\bea\label{phi12}
\phi_1&=&\frac{2v}{\sqrt3}\cos[\pi/3-(1/3)\arccos(j/j_c)]\\
\phi_2&=&\frac{2v}{\sqrt3}\cos[\pi-(1/3)\arccos(j/j_c)]~.\nonumber
\eea

\end{itemize}

We consider below these two situations separately. We note that functional derivatives with respect to a uniform source $j$ are replaced by partial derivatives with respect to $V^{(4)}j$,
and the background field is therefore
\be\label{phib}
\phi_c\equiv\frac{-1}{ZV^{(4)}}\frac{\partial Z}{\partial j}~,
\ee
which is used below.

\subsection{Case $|j|>>j_c$}

This situation describes the region $|\phi_c|\gg v$, where the semi-classical approximation for the partition function is
\be
Z[j]\simeq\exp\Big(-\int d^4x\sqrt{|g|}(U_{bare}(\phi_0)+j\phi_0)\Big)~.
\ee
The background field (\ref{phib}) is then, in this approximation,
\be
\phi_c=(U'_{bare}(\phi_0)+j)\frac{\partial\phi_0}{\partial j}+\phi_0=\phi_0~.
\ee
The semi-classical 1PI effective action is then defined by the Legendre transform
\bea
S_{eff}[\phi_c]&=&-\ln Z[j]-V^{(4)}j\phi_c\\
&=& V^{(4)}U_{bare}(\phi_c)~,
\eea
leading to the semi-classical effective potential 
\be
U_{sc}(\phi_c)= U_{bare}(\phi_c)~.
\ee
The semi-classical approximation therefore does not modify the bare potential for $|\phi_c|\gg v$.

\subsection{Case $|j|<<j_c$}

This situation describes the region $|\phi_c|\ll v$, where the two saddles points $\phi_1,\phi_2$ dominate the partition function. The latter is 
\be
Z[j]\simeq \sum_{k=1}^2 \exp\Big(-\int d^4x\sqrt{|g|}(U_{bare}(\phi_k)+j\phi_k)\Big)~,
\ee
and the background field $\phi_c$ can be obtained as a Taylor expansion in $j$. This expansion can then be inverted to give the source $j$ as a Taylor expansion in $\phi_c$, 
and the 1PI effective action can then be obtained by integrating the equation of motion 
\be
\frac{1}{V^{(4)}}\frac{\partial\Gamma}{\partial\phi_c}=-j~.
\ee
These expansions have been made in \cite{Tsapalis}, and they lead to the semi-classical convex effective action
\bea\label{Gammaexp}
S_{eff}[\phi_c]&=&-K+\frac{4A}{1+8A}\left(\frac{\phi_c}{v}\right)^2\\
&&+\frac{8A(-3+12A+128A^3)}{3(1+8A)^4}\left(\frac{\phi_c}{v}\right)^4\nn
&&+{\cal O}(\phi_c/v)^6~,\nonumber
\eea
where $K$ is a constant and 
\be
A\equiv\frac{\lambda}{24}v^4V^{(4)}~.
\ee

Finally, we note that the generalisation of this semi-classical approximation to $O(N)$ symmetry is also derived in \cite{Tsapalis}, in terms of the radial field component
$r_c=\sqrt{\vec\phi_c\cdot\vec\phi_c}$ and leads to
\be
S_{eff}^N[r_c]=-K_N+a_2\left(\frac{r_c}{v}\right)^2+a_4\left(\frac{r_c}{v}\right)^4+{\cal O}(r_c^6)~,
\ee
where $K_N$ is a constant and
\bea
a_2&=&\frac{4NA}{(1+8A)}\\
a_4&=&\frac{8N^2A[-3N+2(1+5N)A]}{(N+2)(1+8A)^4}\nn
&&~~~~+\frac{8N^2A[32(1-N)A^2+128A^3]}{(N+2)(1+8A)^4}\nn
A&\equiv&\frac{\lambda}{24}V^{(4)}v^4~.\nonumber
\eea


\begin{thebibliography}{99}

\bibitem{Borde:1993xh}
  A.~Borde and A.~Vilenkin,
  %``Eternal inflation and the initial singularity,''
  Phys.\ Rev.\ Lett.\  {\bf 72} (1994) 3305
  doi:10.1103/PhysRevLett.72.3305
  [gr-qc/9312022].
  %%CITATION = doi:10.1103/PhysRevLett.72.3305;%%
  %253 citations counted in INSPIRE as of 30 May 2019

\bibitem{Hawking:1969sw}
  S.~W.~Hawking and R.~Penrose,
  %``The Singularities of gravitational collapse and cosmology,''
  Proc.\ Roy.\ Soc.\ Lond.\ A {\bf 314} (1970) 529.
  doi:10.1098/rspa.1970.0021
  %%CITATION = doi:10.1098/rspa.1970.0021;%%
  %750 citations counted in INSPIRE as of 30 May 2019

\bibitem{barrowandtippler}
    J. D. Barrow and F. J. Tipler. 1985. Mon.Not.Roy.Astron.Soc.,216,395
  %  title = "{Closed universes: their future evolution and final state⋆}",
  %  journal = {Monthly Notices of the Royal Astronomical Society},
  %  volume = {216},
  %  number = {2},
  %  pages = {395-402},
  %  year = {1985},
  %  month = {09},
  %  issn = {0035-8711},
  %  doi = {10.1093/mnras/216.2.395},
  [https://doi.org/10.1093/mnras/216.2.395],
  %  eprint = {http://oup.prod.sis.lan/mnras/article-pdf/216/2/395/3500867/mnras216-0395.pdf},

  
\bibitem{Arnowitt:1960es}
  R.~L.~Arnowitt, S.~Deser and C.~W.~Misner,
  %``Canonical variables for general relativity,''
  Phys.\ Rev.\  {\bf 117} (1960) 1595.
  doi:10.1103/PhysRev.117.1595
  %%CITATION = doi:10.1103/PhysRev.117.1595;%%
  %472 citations counted in INSPIRE as of 27 May 2019

\bibitem{Kleban:2016sqm}
  M.~Kleban and L.~Senatore,
  %``Inhomogeneous Anisotropic Cosmology,''
  JCAP {\bf 1610} (2016) no.10,  022
  doi:10.1088/1475-7516/2016/10/022
  [arXiv:1602.03520 [hep-th]].
  %%CITATION = doi:10.1088/1475-7516/2016/10/022;%%  
  
\bibitem{Steinhardt:2001st}
  P.~J.~Steinhardt and N.~Turok,
  %``Cosmic evolution in a cyclic universe,''
  Phys.\ Rev.\ D {\bf 65} (2002) 126003
  doi:10.1103/PhysRevD.65.126003
  [hep-th/0111098].
  %%CITATION = doi:10.1103/PhysRevD.65.126003;%%
  %633 citations counted in INSPIRE as of 30 May 2019
  
\bibitem{Khoury:2001bz}
  J.~Khoury, B.~A.~Ovrut, N.~Seiberg, P.~J.~Steinhardt and N.~Turok,
  %``From big crunch to big bang,''
  Phys.\ Rev.\ D {\bf 65} (2002) 086007
  doi:10.1103/PhysRevD.65.086007
  [hep-th/0108187].
  %%CITATION = doi:10.1103/PhysRevD.65.086007;%%
  %533 citations counted in INSPIRE as of 30 May 2019
  
\bibitem{Khoury:2001wf}
  J.~Khoury, B.~A.~Ovrut, P.~J.~Steinhardt and N.~Turok,
  %``The Ekpyrotic universe: Colliding branes and the origin of the hot big bang,''
  Phys.\ Rev.\ D {\bf 64} (2001) 123522
  doi:10.1103/PhysRevD.64.123522
  [hep-th/0103239].
  %%CITATION = doi:10.1103/PhysRevD.64.123522;%%
  %1179 citations counted in INSPIRE as of 30 May 2019  
  
%\cite{Guth:1980zm}
\bibitem{Guth:1980zm}
  A.~H.~Guth,
  %``The Inflationary Universe: A Possible Solution to the Horizon and Flatness Problems,''
  Phys.\ Rev.\ D {\bf 23} (1981) 347
   [Adv.\ Ser.\ Astrophys.\ Cosmol.\  {\bf 3} (1987) 139].
  doi:10.1103/PhysRevD.23.347
  %%CITATION = doi:10.1103/PhysRevD.23.347;%%
  %7283 citations counted in INSPIRE as of 30 May 2019

%\cite{Linde:1981mu}
\bibitem{Linde:1981mu}
  A.~D.~Linde,
  %``A New Inflationary Universe Scenario: A Possible Solution of the Horizon, Flatness, Homogeneity, Isotropy and Primordial Monopole Problems,''
  Phys.\ Lett.\  {\bf 108B} (1982) 389
   [Adv.\ Ser.\ Astrophys.\ Cosmol.\  {\bf 3} (1987) 149].
  doi:10.1016/0370-2693(82)91219-9
  %%CITATION = doi:10.1016/0370-2693(82)91219-9;%%
  %4531 citations counted in INSPIRE as of 30 May 2019

%\cite{Albrecht:1982wi}
\bibitem{Albrecht:1982wi}
  A.~Albrecht and P.~J.~Steinhardt,
  %``Cosmology for Grand Unified Theories with Radiatively Induced Symmetry Breaking,''
  Phys.\ Rev.\ Lett.\  {\bf 48} (1982) 1220
   [Adv.\ Ser.\ Astrophys.\ Cosmol.\  {\bf 3} (1987) 158].
  doi:10.1103/PhysRevLett.48.1220
  %%CITATION = doi:10.1103/PhysRevLett.48.1220;%%
  %3954 citations counted in INSPIRE as of 30 May 2019

%\cite{Starobinsky:1980te}
\bibitem{Starobinsky:1980te}
  A.~A.~Starobinsky,
  %``A New Type of Isotropic Cosmological Models Without Singularity,''
  Phys.\ Lett.\ B {\bf 91} (1980) 99
   [Phys.\ Lett.\  {\bf 91B} (1980) 99]
   [Adv.\ Ser.\ Astrophys.\ Cosmol.\  {\bf 3} (1987) 130].
  doi:10.1016/0370-2693(80)90670-X
  %%CITATION = doi:10.1016/0370-2693(80)90670-X;%%
  %4086 citations counted in INSPIRE as of 30 May 2019

%\cite{Brandenberger:2016uzh}
\bibitem{Brandenberger:2016uzh}
  R.~Brandenberger,
  %``Initial conditions for inflation — A short review,''
  Int.\ J.\ Mod.\ Phys.\ D {\bf 26} (2016) no.01,  1740002
  doi:10.1142/S0218271817400028
  [arXiv:1601.01918 [hep-th]].
  %%CITATION = doi:10.1142/S0218271817400028;%%
  %47 citations counted in INSPIRE as of 30 May 2019

\bibitem{East:2015ggf}
  W.~E.~East, M.~Kleban, A.~Linde and L.~Senatore,
  %``Beginning inflation in an inhomogeneous universe,''
  JCAP {\bf 1609} (2016) no.09,  010
  doi:10.1088/1475-7516/2016/09/010
  [arXiv:1511.05143 [hep-th]].
  %%CITATION = doi:10.1088/1475-7516/2016/09/010;%%
  %71 citations counted in INSPIRE as of 30 May 2019

%\cite{Clough:2017efm}
\bibitem{Clough:2017efm}
  K.~Clough, R.~Flauger and E.~A.~Lim,
  %``Robustness of Inflation to Large Tensor Perturbations,''
  JCAP {\bf 1805} (2018) no.05,  065
  doi:10.1088/1475-7516/2018/05/065
  [arXiv:1712.07352 [hep-th]].
  %%CITATION = doi:10.1088/1475-7516/2018/05/065;%%
  %7 citations counted in INSPIRE as of 30 May 2019
  
%\cite{Clough:2016ymm}
\bibitem{Clough:2016ymm}
  K.~Clough, E.~A.~Lim, B.~S.~DiNunno, W.~Fischler, R.~Flauger and S.~Paban,
  %``Robustness of Inflation to Inhomogeneous Initial Conditions,''
  JCAP {\bf 1709} (2017) no.09,  025
  doi:10.1088/1475-7516/2017/09/025
  [arXiv:1608.04408 [hep-th]].
  %%CITATION = doi:10.1088/1475-7516/2017/09/025;%%
  %37 citations counted in INSPIRE as of 30 May 2019
  
\bibitem{Gasperini:1992em}
  M.~Gasperini and G.~Veneziano,
  %``Pre - big bang in string cosmology,''
  Astropart.\ Phys.\  {\bf 1} (1993) 317
  doi:10.1016/0927-6505(93)90017-8
  [hep-th/9211021].
  %%CITATION = doi:10.1016/0927-6505(93)90017-8;%%
  %850 citations counted in INSPIRE as of 07 Jul 2019  

\bibitem{Finelli:2001sr}
  F.~Finelli and R.~Brandenberger,
  %``On the generation of a scale invariant spectrum of adiabatic fluctuations in cosmological models with a contracting phase,''
  Phys.\ Rev.\ D {\bf 65} (2002) 103522
  doi:10.1103/PhysRevD.65.103522
  [hep-th/0112249].
  %%CITATION = doi:10.1103/PhysRevD.65.103522;%%
  %293 citations counted in INSPIRE as of 07 Jul 2019  
  
  
  %\cite{Ijjas:2018qbo}
\bibitem{Ijjas:2018qbo}
  A.~Ijjas and P.~J.~Steinhardt,
  %``Bouncing Cosmology made simple,''
  Class.\ Quant.\ Grav.\  {\bf 35} (2018) no.13,  135004
  doi:10.1088/1361-6382/aac482
  [arXiv:1803.01961 [astro-ph.CO]].
  %%CITATION = doi:10.1088/1361-6382/aac482;%%
  %11 citations counted in INSPIRE as of 30 May 2019

  
\bibitem{Brandenberger:2016vhg}
  R.~Brandenberger and P.~Peter,
  %``Bouncing Cosmologies: Progress and Problems,''
  Found.\ Phys.\  {\bf 47} (2017) no.6,  797
  doi:10.1007/s10701-016-0057-0
  [arXiv:1603.05834 [hep-th]].
  %%CITATION = doi:10.1007/s10701-016-0057-0;%%
  %150 citations counted in INSPIRE as of 07 Jul 2019  

  
\bibitem{bouncemechanism}
  P.~Peter and N.~Pinto-Neto,
  %``Primordial perturbations in a non singular bouncing universe model,''
  Phys.\ Rev.\ D {\bf 66} (2002) 063509
  doi:10.1103/PhysRevD.66.063509
  [hep-th/0203013];
  %%CITATION = doi:10.1103/PhysRevD.66.063509;%%
  %140 citations counted in INSPIRE as of 15 Jun 2019
  T.~Biswas, A.~Mazumdar and W.~Siegel,
  %``Bouncing universes in string-inspired gravity,''
  JCAP {\bf 0603} (2006) 009
  doi:10.1088/1475-7516/2006/03/009
  [hep-th/0508194];
  %%CITATION = doi:10.1088/1475-7516/2006/03/009;%%
  %374 citations counted in INSPIRE as of 15 Jun 2019
  T.~Biswas, R.~Brandenberger, A.~Mazumdar and W.~Siegel,
  %``Non-perturbative Gravity, Hagedorn Bounce & CMB,''
  JCAP {\bf 0712} (2007) 011
  doi:10.1088/1475-7516/2007/12/011
  [hep-th/0610274];
  %%CITATION = doi:10.1088/1475-7516/2007/12/011;%%
  %144 citations counted in INSPIRE as of 15 Jun 2019
  C.~Kounnas, H.~Partouche and N.~Toumbas,
  %``S-brane to thermal non-singular string cosmology,''
  Class.\ Quant.\ Grav.\  {\bf 29} (2012) 095014
  doi:10.1088/0264-9381/29/9/095014
  [arXiv:1111.5816 [hep-th]].
  %%CITATION = doi:10.1088/0264-9381/29/9/095014;%%
  %27 citations counted in INSPIRE as of 15 Jun 2019

  
\bibitem{Tsapalis} 
  J.~Alexandre and A.~Tsapalis,
  %``Maxwell Construction for Scalar Field Theories with Spontaneous Symmetry Breaking,''
  Phys.\ Rev.\ D {\bf 87}, no. 2, 025028 (2013)
  doi:10.1103/PhysRevD.87.025028
  [arXiv:1211.0921 [hep-th]].
  %%CITATION = doi:10.1103/PhysRevD.87.025028;%%
  %10 citations counted in INSPIRE as of 14 May 2019  
    

\bibitem{Miransky} 
  V.~A.~Miransky,
  %``Dynamical symmetry breaking in quantum field theories,''
  Singapore: World Scientific (1993) 533 p
  %11 citations counted in INSPIRE as of 14 May 2019
    
\bibitem{Plascencia:2015pga} 
  A.~D.~Plascencia and C.~Tamarit,
  %``Convexity, gauge-dependence and tunneling rates,''
  JHEP {\bf 1610}, 099 (2016)
  doi:10.1007/JHEP10(2016)099
  [arXiv:1510.07613 [hep-ph]].
  %%CITATION = doi:10.1007/JHEP10(2016)099;%%
  %32 citations counted in INSPIRE as of 14 May 2019    
    

\bibitem{convexity} 
  K.~Symanzik,
  %``Renormalizable models with simple symmetry breaking. 1. Symmetry breaking by a source term,''
  Commun.\ Math.\ Phys.\  {\bf 16}, 48 (1970).
  doi:10.1007/BF01645494;
  %%CITATION = doi:10.1007/BF01645494;%%
  %301 citations counted in INSPIRE as of 14 May 2019
  J.~Iliopoulos, C.~Itzykson and A.~Martin,
  %``Functional Methods and Perturbation Theory,''
  Rev.\ Mod.\ Phys.\  {\bf 47}, 165 (1975).
  doi:10.1103/RevModPhys.47.165;
  %%CITATION = doi:10.1103/RevModPhys.47.165;%%
  %361 citations counted in INSPIRE as of 14 May 2019
  R.~W.~Haymaker and J.~Perez-Mercader,
  %``Convexity of the Effective Potential,''
  Phys.\ Rev.\ D {\bf 27}, 1948 (1983).
  doi:10.1103/PhysRevD.27.1948
  %%CITATION = doi:10.1103/PhysRevD.27.1948;%%
  %43 citations counted in INSPIRE as of 14 May 2019
  
  
\bibitem{competition} 
  S.~R.~Coleman, R.~Jackiw and H.~D.~Politzer,
  %``Spontaneous Symmetry Breaking in the O(N) Model for Large N*,''
  Phys.\ Rev.\ D {\bf 10}, 2491 (1974).
  doi:10.1103/PhysRevD.10.2491;
  %%CITATION = doi:10.1103/PhysRevD.10.2491;%%
  %487 citations counted in INSPIRE as of 14 May 2019
  Y.~Fujimoto, L.~O'Raifeartaigh and G.~Parravicini,
  %``Effective Potential for Nonconvex Potentials,''
  Nucl.\ Phys.\ B {\bf 212}, 268 (1983);
  doi:10.1016/0550-3213(83)90305-X
  %%CITATION = doi:10.1016/0550-3213(83)90305-X;%%
  R.~J.~Rivers,
  %``Effective Potential Convexity and Finite Temperature Phase Transitions,''
  Z.\ Phys.\ C {\bf 22} (1984) 137.
  doi:10.1007/BF01572161
  %%CITATION = doi:10.1007/BF01572161;%%
  %33 citations counted in INSPIRE as of 06 Jul 2132 citations counted in INSPIRE as of 14 May 2019  
  


\bibitem{Wilsonian}
  C.~Wetterich,
  %``Average Action and the Renormalization Group Equations,''
  Nucl.\ Phys.\ B {\bf 352} (1991) 529;
  %%doi:10.1016/0550-3213(91)90099-J
  %%CITATION = doi:10.1016/0550-3213(91)90099-J;%%
  %271 citations counted in INSPIRE as of 20 Jun 2016
  J.~Alexandre, V.~Branchina and J.~Polonyi,
  %``Instability induced renormalization,''
  Phys.\ Lett.\ B {\bf 445}, 351 (1999)
  doi:10.1016/S0370-2693(98)01491-9
  [cond-mat/9803007].
  %%CITATION = doi:10.1016/S0370-2693(98)01491-9;%%
  %79 citations counted in INSPIRE as of 14 May 2019
  
  
\bibitem{Maxwell}
C.~M.~Bender and F.~Cooper,
  %``Failure of the Naive Loop Expansion for the Effective Potential in $\phi^4$ Field Theory When There Is 'Broken Symmetry',''
  Nucl.\ Phys.\ B {\bf 224}, 403 (1983).
  doi:10.1016/0550-3213(83)90383-8;
  %%CITATION = doi:10.1016/0550-3213(83)90383-8;%%
  %44 citations counted in INSPIRE as of 24 May 2019
  F.~Cooper and B.~Freedman,
  %``Renormalizing the Effective Potential for Spontaneously Broken $g \phi^4$ Field Theory,''
  Nucl.\ Phys.\ B {\bf 239}, 459 (1984).
  doi:10.1016/0550-3213(84)90258-X;
  %%CITATION = doi:10.1016/0550-3213(84)90258-X;%%
  %26 citations counted in INSPIRE as of 24 May 2019  
  P.~Millington and P.~M.~Saffin,
  %``Visualising quantum effective action calculations in zero dimensions,''
  arXiv:1905.09674 [hep-th].
  %%CITATION = ARXIV:1905.09674;%%    
  
  
\bibitem{spinodal}  
  D.~Boyanovsky, H.~J.~de Vega, R.~Holman and J.~F.~J.~Salgado,
  %``Analytic and numerical study of preheating dynamics,''
  Phys.\ Rev.\ D {\bf 54}, 7570 (1996)
  doi:10.1103/PhysRevD.54.7570
  [hep-ph/9608205];
  %%CITATION = doi:10.1103/PhysRevD.54.7570;%%
  %223 citations counted in INSPIRE as of 14 May 2019
  D.~Boyanovsky, D.~Cormier, H.~J.~de Vega, R.~Holman, A.~Singh and M.~Srednicki,
  %``Preheating in FRW universes,''
  hep-ph/9609527;
  %%CITATION = HEP-PH/9609527;%%
  %24 citations counted in INSPIRE as of 14 May 2019
  D.~Cormier and R.~Holman,
  %``Spinodal decomposition and inflation: Dynamics and metric perturbations,''
  Phys.\ Rev.\ D {\bf 62}, 023520 (2000)
  doi:10.1103/PhysRevD.62.023520
  [hep-ph/9912483];
  %%CITATION = doi:10.1103/PhysRevD.62.023520;%%
  %37 citations counted in INSPIRE as of 14 May 2019
   S.~Tsujikawa and T.~Torii,
  %``Spinodal effect in the natural inflation model,''
  Phys.\ Rev.\ D {\bf 62}, 043505 (2000)
  doi:10.1103/PhysRevD.62.043505
  [hep-ph/9912499].
  %%CITATION = doi:10.1103/PhysRevD.62.043505;%%
  %13 citations counted in INSPIRE as of 14 May 2019 
  
\bibitem{Alex} 
  J.~Alexandre,
  %``Convexity at finite temperature and non-extensive thermodynamics,''
  Nucl.\ Phys.\ B {\bf 910}, 868 (2016)
  doi:10.1016/j.nuclphysb.2016.07.030
  [arXiv:1603.01540 [hep-th]].
  %%CITATION = doi:10.1016/j.nuclphysb.2016.07.030;%%
  %1 citations counted in INSPIRE as of 14 May 2019    
 
\bibitem{Zeldovich:1984vk}
  Y.~B.~Zeldovich and A.~A.~Starobinsky,
  %``Quantum creation of a universe in a nontrivial topology,''
  Sov.\ Astron.\ Lett.\  {\bf 10} (1984) 135.
  %%CITATION = SALED,10,135;%%
  %222 citations counted in INSPIRE as of 29 Sep 2019
  
\bibitem{reviewCasimir}
  M.~Bordag, U.~Mohideen and V.~M.~Mostepanenko,
  %``New developments in the Casimir effect,''
  Phys.\ Rept.\  {\bf 353} (2001) 1
  doi:10.1016/S0370-1573(01)00015-1
  [quant-ph/0106045].
  %%CITATION = doi:10.1016/S0370-1573(01)00015-1;%%
  %841 citations counted in INSPIRE as of 29 Sep 2019
 
\bibitem{Herdeiro:2005zj}
  C.~A.~R.~Herdeiro and M.~Sampaio,
  %``Casimir energy and a cosmological bounce,''
  Class.\ Quant.\ Grav.\  {\bf 23} (2006) 473
  doi:10.1088/0264-9381/23/2/012
  [hep-th/0510052].
  %%CITATION = doi:10.1088/0264-9381/23/2/012;%%
  %31 citations counted in INSPIRE as of 29 Sep 2019 
 
\bibitem{DiTucci:2019xcr}
  A.~Di Tucci, J.~Feldbrugge, J.~L.~Lehners and N.~Turok,
  %``Quantum Incompleteness of Inflation,''
  arXiv:1906.09007 [hep-th].
  %%CITATION = ARXIV:1906.09007;%%  

\bibitem{Bramberger:2019zks}
  S.~F.~Bramberger, A.~Di Tucci and J.~L.~Lehners,
  %``Homogeneous Transitions during Inflation: a Description in Quantum Cosmology,''
  arXiv:1907.05782 [gr-qc].  
    
\bibitem{bounce} 
  S.~R.~Coleman,
  %``The Fate of the False Vacuum. 1. Semiclassical Theory,''
  Phys.\ Rev.\ D {\bf 15}, 2929 (1977)
  Erratum: [Phys.\ Rev.\ D {\bf 16}, 1248 (1977)].
  doi:10.1103/PhysRevD.15.2929, 10.1103/PhysRevD.16.1248;
  %%CITATION = doi:10.1103/PhysRevD.15.2929, 10.1103/PhysRevD.16.1248;%%
  %1890 citations counted in INSPIRE as of 14 May 2019
  C.~G.~Callan, Jr. and S.~R.~Coleman,
  %``The Fate of the False Vacuum. 2. First Quantum Corrections,''
  Phys.\ Rev.\ D {\bf 16}, 1762 (1977).
  doi:10.1103/PhysRevD.16.1762
  %%CITATION = doi:10.1103/PhysRevD.16.1762;%%
  %1207 citations counted in INSPIRE as of 14 May 2019
 
  
\end{thebibliography}
\end{document}